\documentclass[11pt]{article}
\usepackage[textwidth=15.2cm,textheight=22cm]{geometry}
\usepackage{amsmath,amssymb}
\usepackage{latexsym}
\usepackage{multicol}
\usepackage{graphicx}
\usepackage{bm}
\usepackage{varwidth}
\usepackage{pifont}
\usepackage{cancel}
\usepackage{xcolor}
\usepackage{tikz-feynman,contour} %New package for feynman diagrams+label
\usepackage{subcaption}
\usepackage[title]{appendix}
\usepackage{cite}

\allowdisplaybreaks[1]

\newcommand{\del}{\partial}
\newcommand{\be}{\begin{equation}}
\newcommand{\ee}{\end{equation}}
\newcommand{\ba}{\begin{eqnarray}}
\newcommand{\ea}{\end{eqnarray}}
\newcommand{\bdm}{\begin{displaymath}}
\newcommand{\edm}{\end{displaymath}}

\newcommand{\rom}[1]{\uppercase\expandafter{\romannumeral #1\relax}}

\def\ba{\bar A}

\def\beq{\begin{equation}}
\def\eeq{\end{equation}}

\newcommand{\nn}{\nonumber}

\newcommand{\ndt}{\noindent}

\def\bea{\begin{eqnarray}}
\def\eea{\end{eqnarray}}
\def\beas{\begin{eqnarray*}}
\def\eeas{\end{eqnarray*}}
\def\sla{\raise.15ex\hbox{$/$}\kern-.57em}

\def\spa#1.#2{\left\langle#1\,#2\right\rangle}
\def\spb#1.#2{\left[#1\,#2\right]}

\begin{document}

\begin{titlepage}
\begin{flushright}    
{\small $\,$}
\end{flushright}
\vskip 1cm
\centerline{\Large{\bf{Soft factors and interaction vertices from light-cone actions}}}
\vskip 1cm
\centerline{Sudarshan Ananth, Chetan Pandey and Saurabh Pant}
\vskip 0.3cm
\centerline{\it {Indian Institute of Science Education and Research}}
\centerline{\it {Pune 411008, India}}
\vskip 1.5cm
\centerline{\bf {Abstract}}
\vskip .5cm
\ndt Universal factors associated with the emission of a soft boson in gauge theories and gravity, formulated in the light-cone gauge, are presented.  The inverse-soft method, for constructing higher-point amplitudes from lower-point ones, using these factors is reviewed. These ideas are then examined in (light-cone) superspace and applied to both the $\mathcal{N}=4$ super Yang-Mills and $\mathcal{N}=8$ supergravity theories. One highlight is a compact result for the quartic interaction vertex in $\mathcal{N}=8$ supergravity, a crucial ingredient for finiteness analyses. 
\vfill
\end{titlepage}

\section{Introduction}

\ndt The light-cone gauge offers an interesting perspective on scattering amplitudes. Inherently non-local and non-covariant, it focuses exclusively on the physical degrees of freedom in the theory, rendering the `physics' manifest. Spurious degrees of freedom and redundancies do not obscure the symmetries in a theory, and the compact spinor helicity variables emerge naturally in this gauge. Light-cone actions, unlike most covariant actions, may be explicitly derived from first principles - a feature that follows from the requirement of closure of the Poincar\' e algebra of which the light-cone Hamiltonian is itself an element~\cite{BBB1}.
\vskip 0.3cm
 \ndt This algebra-closure approach allows us to obtain consistent interaction vertices order by order in the coupling constant. In the light-cone gauge, where we work in a helicity basis, interaction vertices are closely linked with scattering amplitudes. A somewhat complementary approach to derive higher-point amplitudes directly is to construct them from lower-point ones by using a multiplicative universal factor, associated with the emission of a soft boson~\cite{Arkani-Hamed:2009ljj}.
\vskip 0.3 cm
\ndt In this paper, we adapt the inverse-soft method to the light-cone gauge and review the procedure for gauge theories and gravity. We then move to light-cone superspace and apply these results to the maximally supersymmetric $\mathcal{N}=4$ super Yang-Mills and $\mathcal{N}=8$ supergravity theories. In this process, we obtain a compact expression for the quartic interaction vertex in $\mathcal N=8$ supergravity, an essential ingredient for both an $\mathcal{N}=8$ MHV Lagrangian~\cite{L4} and a finiteness analysis~\cite{BLN2,SM,Ananth:2006ac} of the theory.

\vskip 0.5cm

\section{Inverse soft method for gauge theories and gravity}
\ndt We start with a $(n\!+\!1)$-particle amplitude $M_{n+1}(p_1, p_2,\cdots, p_n, k)$ and consider the limit in which the momentum of the external particle, with momentum $k$, vanishes. The amplitude then, at leading order, factorizes~\cite{Weinberg:1965nx} as

\bea
\lim_{k\to 0}\,M_{n+1}(p_1, p_2,\cdots, k)=S\times M_n(p_1, p_2, \cdots p_n)\,+\,\mathcal O(k^0)\ ,
\eea
where $S$ is the universal soft factor associated with the emission of the momentum-$k$ particle.
\vskip 0.7cm

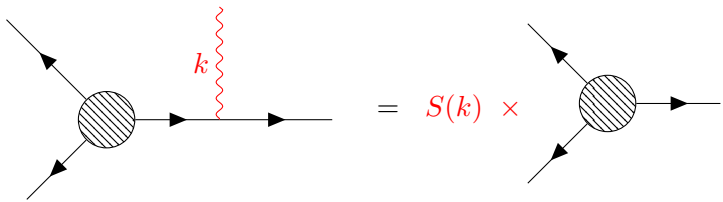
\begin{tabular}{ccc}
\begin{tabular}{c}
\begin{tikzpicture}
\begin{feynman}
\vertex (a);
\vertex [blob, below right=of a] (b) {};
\vertex [right=of b] (c);
\vertex [right=of c] (d);
\vertex [above=of c] (e);
\vertex [below left=of b] (f);

\diagram* {
    (b) -- [fermion] (a),
    (b) -- [fermion] (c) -- [fermion] (d),
    (c) -- [photon, red, edge label=$k$] (e),
    (b) -- [fermion] (f),
};
\end{feynman}
\end{tikzpicture}
\end{tabular}
&
\begin{tabular}{c}
\!\!\!\!=\color{red}{$\;\;\ S(k)\;\;\times$}
\end{tabular}
&
\!\!\!\!\!\!\!\!\!\!\!\!\!\!
\begin{tabular}{c}
\begin{tikzpicture}
\begin{feynman}

\vertex [blob, right=2cm of d] (k) {};
\vertex [above left=of k] (j);
\vertex [right=of k] (l);
\vertex [below left=of k] (m);

\diagram* {
    (k) -- [fermion] (j),
    (k) -- [fermion] (l),
    (k) -- [fermion] (m),
};
\end{feynman}
\end{tikzpicture}
\end{tabular}
\end{tabular}
\\

\ndt As a consequence, a $(n\!+\!1)$-particle amplitude may be obtained by multiplying a $n$-particle amplitude with the universal soft factor - the inverse soft approach. This is a valuable method because the cubic vertices, in the light-cone formalism, are genuinely easy to derive. We use light-cone coordinates 
\begin{eqnarray}
x^{\pm}=\frac{x^{0}\pm x^{3}}{\sqrt{2}} \;,\qquad
x = \frac{x^{1}+ix^{2}}{\sqrt{2}} \;,\qquad \bar x= x^*\ .
\end{eqnarray}
and their derivatives $\partial_\pm, \bar\partial,\partial$. The metric is $(-,+,+,+)$, $x^+$ plays the role of time and we define the operator $\frac{1}{\partial^+}$ following the prescription in~\cite{SM}. The off-shell spinor products in this language are
\bea
\label{sh}
\langle kl\rangle \equiv \sqrt{2}\,\frac{(kl^{+}-lk^{+})}{\sqrt{k^{+}l^{+}}},\hspace{1cm}  [ kl] \equiv \sqrt{2}\,	\frac{(\bar kl^{+}-\bar lk^{+})}{\sqrt{k^{+}l^{+}}}\ .
\eea
\ndt Their product is
\bea
\langle kl\rangle[lk]=-(k+l)^2\,.
\eea
\vskip 0.5cm

\subsection{Gauge theory}

\ndt Consider a general Feynman diagram with a number of external scalar legs, with one of these legs emitting an external photon carrying a momentum $k$. In the `soft photon' limit, to leading order, the amplitude corresponding to this diagram $M_{n+1}(p_1,p_2\cdots p_n,k)$ factorizes into an amplitude sans the external photon $M_n(p_1,p_2\cdots,p_n)$ times a (scalar-scalar-photon) cubic vertex.
\vskip 0.5cm

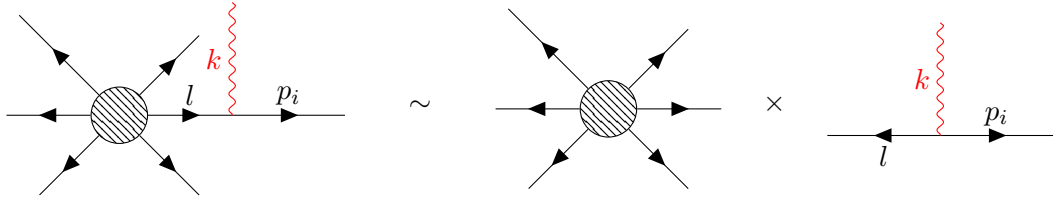
\begin{tabular}{ccc}
\begin{tabular}{c}
\begin{tikzpicture}
\begin{feynman}
\vertex (a);
\vertex [blob, below right=of a] (b) {};
\vertex [right=of b] (c);
\vertex [left=of b] (g);
\vertex [above right=of b] (h);
\vertex [below right=of b] (i);
\vertex [right=of c] (d);
\vertex [above=of c] (e);
\vertex [below left=of b] (f);

\diagram* {
    (b) -- [fermion] (a),
    (b) -- [fermion, edge label=$l$] (c) -- [fermion, edge label=$p_i$] (d),
    (c) -- [photon, red, edge label=$k$] (e),
    (b) -- [fermion] (f),
    (b) -- [fermion] (g),
    (b) -- [fermion] (h),
    (b) -- [fermion] (i),
};
\end{feynman}
\end{tikzpicture}
\end{tabular}
&
\begin{tabular}{c}
$\sim$
\end{tabular}
&
\begin{tabular}{c}
\begin{tikzpicture}
\begin{feynman}
\vertex (a);
\vertex [blob, below right=of a] (b) {};
\vertex [right=of b] (c);
\vertex [left=of b] (g);
\vertex [above right=of b] (h);
\vertex [below right=of b] (i);
\vertex [below left=of b] (f);

\diagram* {
    (b) -- [fermion] (a),
    (b) -- [fermion] (c),
    (b) -- [fermion] (f),
    (b) -- [fermion] (g),
    (b) -- [fermion] (h),
    (b) -- [fermion] (i),
};
\end{feynman}
\end{tikzpicture}
\end{tabular}
\begin{tabular}{c}
$\times$
\end{tabular}
\begin{tabular}{c}
\begin{tikzpicture}
\begin{feynman}
\vertex (a);
\vertex [right= of a ] (b);
\vertex [right= of b] (c);
\vertex [above= of b] (d);

\diagram* {
    (b) -- [fermion, edge label=$l$] (a),
    (b) -- [fermion, edge label=$p_i$] (c),
    (b) -- [photon, red, edge label=$k$] (d),
};
\end{feynman}
\end{tikzpicture}
\end{tabular}
\end{tabular}

\ndt where $p_i$ denotes the $i^{\text{th}}$ external leg in the diagram.

\vskip 0.3cm

\ndt The relevant action here couples a complex scalar field to a photon~\cite{Akshay:2014qea}.
\bea
S=\int d^4x\left[\bar A\,\square \, A+\bar\phi\,\square\,\phi +g\left(A\;\bar\phi\;\bar\partial\,\phi-\frac{\bar\partial}{\partial^+}\,A\;\bar\phi\;\partial^+\,\phi\right)\right]\ ,
\eea
\ndt with the momentum-space cubic Lagrangian
\bea
\left(\bar p_i-p_i^+\frac{\bar k}{k^+}\right)\tilde{A}(k)\;\tilde{\bar\phi}(l)\;\tilde{\phi}(p_i)\;\delta^4(p_i+l+k)\ .
\eea
For the amplitude, this means
\bea
M_{n+1}(p_i,k) = \bigg\{\,g\,\sum_{i=1}^n\frac{1}{(p_i\pm k)^2}\left(\bar p_i-p_i^+\frac{\bar k}{k^+}\right)\delta^4(p_i+l+k) \bigg\} M_n(p_i -k) \,,
\eea
\ndt where we sum over the particle emission from all external legs. \vspace{3 mm} \\
\ndt In the soft limit, the amplitude factorizes as
\bea
M_{n+1}&\sim& g\sum_{i=1}^n\frac{1}{(p_i\cdot k)}\left(\frac{[ p_i k]\sqrt{p_i^+k^+}}{k^+}\right)\delta^4(p_i+l+k)\, M_n\,, \nn \\
&\sim&g\sum_{i=1}^n\frac{1}{\langle p_i k\rangle}\left(\sqrt{\frac{p_i^+}{k^+}}\right)\delta^4(p_i+l+k)\, M_n\,.
\eea

\ndt From this, we read of the soft factor for photon emission
\bea
s_P=\sum_{i=1}^n\frac{1}{\langle p_i k\rangle}\sqrt{\frac{p_i^+}{k^+}}=\sum_{i=1}^n\frac{\langle p_il\rangle}{\langle p_ik\rangle \langle kl\rangle}\,.
\eea

\ndt In Yang-Mills theory at tree level, due to color ordering of the amplitudes, the soft factor only depends on the legs adjacent to the soft particle.
\bea
\label{sft}
s_{YM}=\frac{\langle n,1\rangle}{\langle n,n+1\rangle\langle n+1,1 \rangle} \,.
\eea

\ndt We note that there is an alternate approach to obtaining (\ref {sft}). `Make' a gluon soft in the four-gluon amplitude, which then factorizes into the three-gluon amplitude times the soft factor.
\bea
 M_4^{YM}=\frac{\langle 12\rangle^4}{\langle 12\rangle\langle 23\rangle\langle 34\rangle\langle 41\rangle}\Rightarrow \lim_{4\to 0}  M_4^{YM}=\underbrace{\frac{\langle 12\rangle^4}{\langle 12\rangle\langle 23\rangle\langle 31\rangle}}_{ M_3^{YM}}\times \color{red}{\underbrace{\frac{\langle 31\rangle}{\langle 34\rangle\langle 41\rangle}}_{s_{YM}}}\,.
\eea

\ndt All higher-point amplitudes can be constructed by taking a product of lower-point amplitude times a universal soft factor. For gauge theories, the inverse-soft recursion relation is
\bea
M_{n+1}(p_1, p_2,\cdots,p_n, k)=S\times M_n(p_1', p_2, \cdots p_n')\,,
\eea
\ndt To conserve momentum on the right side, the momentum of adjacent particles $p_1$ and $p_n$ must be shifted; the prime indicates this. \\

\ndt We note at this point that we can obtain `off-shell' MHV vertices by defining appropriate off-shell spinor helicity variables and taking all the external legs in an MHV amplitude off-shell \cite{Cachazo:2004kj}. The $(n+1)$-point MHV vertex then reads
\bea
V_{n+1}(1^-, 2^-,3^+,\cdots,n^+, k^+) &=&\frac{\langle 12\rangle^4}{\langle 12\rangle\langle 23\rangle\cdots\langle n 1\rangle} \times \frac{\langle n 1\rangle}{\langle nk \rangle\langle k1\rangle}\,, \nn\\
&=& \frac{\langle 12\rangle^4}{\langle 12\rangle\langle 23\rangle\cdots\langle n k \rangle\langle k1\rangle}\,.
\eea
In the specific case of Yang-Mills theory, the entire Lagrangian, in the light-cone gauge, may be mapped to a MHV Lagrangian \cite{L1,L2}.

\subsection{Gravity}

\ndt The Lagrangian in momentum-space, at cubic order in the light-cone gauge, for a scalar coupled to a graviton is~\cite{Akshay:2014qea}
\bea
\mathcal{L}_{\,G}^3=\left(\bar p_i^2-2p_i^+\bar p_i\frac{\bar k}{k^+}+p_i^{+2}\frac{\bar k^2}{k^{+2}}\right)\tilde{h}(k)\tilde{\bar\phi}(r)\tilde{\phi}(p_i)\delta^4(p_i+r+k)\ .
\eea
\ndt In the soft graviton limit, the $(n\!+\!1)$-point amplitude factorizes as
\bea
M_{n+1}\!\!\!\!\!\!\!\!\!&&\approx M_n\;g\sum_{i=1}^n\frac{1}{\langle p_ik\rangle}\left(\frac{[p_i k]\langle p_i r\rangle^2}{\langle r k\rangle^2}\right)\delta^4(p_i+r+k)\,.
\eea
\ndt The soft factor for graviton emission then reads
\bea
\label{gravity}
s_{G}=\sum_{i=1}^n\frac{[p_i k]}{\langle p_i k\rangle}\frac{\langle p_i r\rangle^2}{\langle r k\rangle^2}=\sum_{i=1}^n\frac{[p_i k]}{\langle p_i k\rangle}\frac{\langle p_ir\rangle\langle sp_i\rangle}{\langle rk\rangle\langle sk\rangle}\,,
\eea
\ndt where $r$ and $s$ are arbitrary vectors for reference legs. This follows from
\bea
\sum_{i=1}^n\frac{[p_i k]}{\langle p_i k\rangle}\frac{\langle p_i r\rangle^2}{\langle r k\rangle^2}=\sum_{i=1}^n\frac{[p_i k]}{\langle p_i k\rangle}\frac{\langle p_i r\rangle\langle p_ir\rangle\langle sk\rangle}{\langle kr\rangle\langle kr\rangle\langle sk\rangle}=\sum_{i=1}^n\frac{[p_i k]}{\langle p_i k\rangle}\frac{\langle p_ir\rangle\langle sp_i\rangle}{\langle rk\rangle\langle sk\rangle}+\underbrace{\sum_{i=1}^n\frac{[p_i k]}{\langle p_i k\rangle}\frac{\langle p_i r\rangle\langle kp_i\rangle\langle rs\rangle}{\langle kr\rangle\langle kr\rangle\langle sk\rangle}}_{=0}\,, \nn
\eea

\ndt where we have used the Schouten identity in the last equality, and the final expression vanishes upon imposing momentum conservation $\sum_i |p_i\rangle[p_i|=0$.\\

\ndt To apply the inverse-soft method here, we start with the three-point MHV gravity amplitude and multiply it by the universal soft factor so we have
\bea
V_4^G=\left[\frac{\langle pl\rangle^8}{\langle qp\rangle^2\langle pl\rangle^2\langle lq\rangle^2}\right]\times \frac{[p k]}{\langle p k\rangle}\frac{\langle p l\rangle\langle p q\rangle}{\langle l k\rangle\langle k q\rangle}=\frac{\langle lp\rangle^8[pk]}{\langle kp\rangle\langle kq\rangle\langle ql\rangle^2\langle pl\rangle \langle lk\rangle\langle pq\rangle}\,,
\eea
\\
\ndt which is the four-graviton vertex in~\cite{L3}. The word `vertex' here is justified because the off-shell structure at this order, in the light-cone gauge, takes the form~\cite{L3} 
\bea
\mathcal{L}_{\,G}^4 = \kappa^2\,\left(\; V_4^G (p,l,q,k)+ \, f(p,l,q,k)\,p^2 \;\right)\, \bar{h}(p)\bar{h}(l) {h}(q){h}(k)\,.
\eea
The $f$ term above may be eliminated at this order, by a suitable field redefinition~\cite{L3,BCL,ABHS}. Essentially, in this framework, the interaction vertex and the corresponding amplitude are equivalent as long as the external legs are viewed as being off-shell.

\vskip 0.5cm
\subsection{Universality}
\vskip 0.3cm

\ndt The soft factor is universal, i.e., independent of the spins of the hard particles, allowing us to consider the simplest case of scalar external legs. We could have, equivalently, considered a more involved vertex, for example of the type spin $\!1$-spin $\!1$-spin $\!1$ where the hard particles have spin $1$ as well. To see this, start from equation (24) in \cite{Akshay:2014qea} to obtain
\bea
 M_{n+1}\!\!\!\!\!\!\!\!&&\approx M_n\,\frac{gq_i}{(p\pm k)^2}\left(\bar p-p^+\frac{\bar k}{k^+}\right)\times\frac{p^++k^+}{p^+}\delta^4(p+l+k)\,, \nn\\
&& \sim M_n\,\underbrace{\frac{gq_i}{(p\pm k)^2}\left(\bar p-p^+\frac{\bar k}{k^+}\right)}_{S_{YM}}\times\left(1+\frac{k^+}{p^+}\right)\delta^4(p+l+k)\,.
\eea

\ndt We observe that the leading order term is the universal soft factor for the emitted gluon, while the sub-leading term is reminiscent of the superrotation generator in light-cone language \cite{Cachazo:2014fwa}. A similar calculation with the hard particles having spin $2$ (with soft gluon emission) gives us
\bea
 M_{n+1}\!\!\!\!\!\!\!\!&&\approx M_n\,\frac{gq_i}{(p\pm k)^2}\left(\bar p-p^+\frac{\bar k}{k^+}\right)\times\frac{(p^++k^+)^2}{p^{+2}}\delta^4(p+l+k)\,, \nn \\
&& \sim M_n\,\underbrace{\frac{gq_i}{(p\pm k)^2}\left(\bar p-p^+\frac{\bar k}{k^+}\right)}_{S_{YM}}\times\left(1+\frac{k^+}{p^+}\right)^2\delta^4(p+l+k)\, ,
\eea
\ndt confirming that the soft factor is independent of the spins of the hard particles (although sub-leading and higher order terms may depend on these \cite{Cachazo:2014fwa}).
\vskip 0.3cm

\section{$\mathcal{N}=4$ super Yang-Mills theory in light-cone superspace}
\ndt We now move to examining the soft theorem and its consequences in light-cone superspace. We start by discussing the $\mathcal{N}=4$ Yang-Mills theory in four dimensions. This theory involves a complex bosonic field (the gauge field), four complex Grassmann fields, and six scalar fields. Its Lagrangian may be derived in two ways. The first method is to impose the light-cone gauge choice and then eliminate unphysical fields~\cite{BLN1}. In the second method, discussed already, one starts with a superfield and requires closure of the  super Poincar\'e algebra, resulting in a consistent Hamiltonian~\cite{BBB2}.
\vskip 0.3cm
\ndt We introduce anticommuting Grassmann variables $\theta^m$ and $\bar{\theta}_m$, 
\bea
 \{\,\theta^m \;,\; \theta^n \,\} \;\;=\;\; \{\,\bar{\theta}_m \;,\; \bar{\theta}_n \,\} \;\;=\;\; \{\,\bar{\theta}_m \;,\;{\theta^n} \,\} =0 \,,
\eea
where $m,n,p,q,...=1,2,3,4,$ denote $SU(4)$ spinor indices. \\

\ndt All physical degrees of freedom are captured by a single complex superfield
\bea
\label{orig}
\phi(y)&=&\frac{1}{\del^+} A(y) + \frac{i}{\sqrt{2}} \theta^m \theta^n \bar{C}_{mn}(y) + \frac{1}{12} \theta^m\theta^n\theta^p\theta^q \epsilon_{mnpq} \del^+ \bar{A}_(y) \nn\\
&&+ \frac{i}{\del^+} \theta^m \bar{\chi}_m (y) + \frac{\sqrt{2}}{6} \theta^m\theta^n\theta^p \epsilon_{mnpq} \chi^q (y)\,, 
\eea
where $y=(\,x,\bar{x},x^+, y^- \equiv x^- - \frac{i}{\sqrt{2}}\theta^m \bar{\theta}_m \,)$. The gauge fields are $A$ and $\bar{A}$, $\chi^m$ and $\bar{\chi}_m$ are the fermions and the six scalar fields are $\bar{C}_{mn}$ expressed as antisymmetric $SU(4)$ bi-spinors. \\

\ndt The chiral derivatives 
\bea
d^m = - \del^m -\frac{i}{\sqrt{2}}\,\theta^m \del^+\; ; \hspace{1cm} \bar{d}_n =\bar{\partial}_n +  \frac{i}{\sqrt{2}}\,\bar{\theta}_n \del^+\,,
\eea
satisfy 
\bea
 \{\, d^m \;,\; \bar{d}_n\,\}\;=\;- i\sqrt{2}\, \delta^m_{\;n}\, \del^+ \,.
\eea
\ndt The superfield $\phi$ and its complex conjugate $\bar{\phi}$ satisfy the chiral constraints
\bea
d^m\,\phi=0\;; \hspace{1 cm} \bar{d}_n\,\bar{\phi}=0 \,,
\eea
as well as the `inside-out' constraints
\bea
\label{ior}
\bar{\phi}(x, \theta,\bar{\theta} ) = \frac{1}{48}\, \frac{\bar{d}^4}{\del^{+\,2}} \,\phi(x,  \theta,\bar{\theta} ) \,,
\eea
where $\bar{d}^4 = \epsilon^{mnpq} \bar{d}_m  \bar{d}_n \bar{d}_p \bar{d}_q$\,.
\vskip 1 cm
\subsection{Soft factor}
\ndt The super Yang-Mills action is 
\bea
\label{Sym}
S = \int\! d^4x\; d^4\theta\; d^4\bar{\theta}\; \bigg\{\; \frac{1}{4} \,\bar{\phi}^a \frac{\Box}{\del^{+\,2}}\, \phi^a + \frac{2}{3}\,g f^{abc}\, \left(\frac{1}{\del^{+\,2}}\,\bar{\phi}^a\, \bar{\del} \phi^b\, \del^+ \phi^c + c.c \right) + \mathcal{O}(g^2) \,\bigg\}\,,
\eea
with Grassmann integration normalized so $\int\! d^4 \theta\,\theta^4 = 1$. \\

\ndt The momentum space cubic Lagrangian reads (measure and constants suppressed)
\bea
\label{Cubic}
gf^{abc}\, \frac{1}{l^{+\,2}} \, (\,\bar{p}_i k^+ -\bar{k} p_i^+\, ) \,\tilde{\bar{\phi}}^a(l) \tilde{\phi}^b(p_i) \tilde{\phi}^c(k) \,\delta^4(p_i+k+l)\,.
\eea

\ndt In the soft limit $k\rightarrow 0$, the superamplitude $\mathcal{M}_{n+1}$ factorizes as
\bea
\mathcal{M}_{n+1}(p_1,p_2\cdots p_n,k) \sim \bigg\{\,g\sum_{i=p,q} \, \frac{ 1}{l^2 } \, (\,\bar{p}_i k^+ -\bar{k} p^+_i\, )\,\delta^4(p_i+k+l)\bigg\} \mathcal{M}_n(p_1,p_2\cdots,p_n) \,,
\eea
\ndt with a sum over the particle emission from all possible external legs. Invoking momentum conservation, we obtain
\bea
\mathcal{M}_{n+1}&\sim&  g\,\sum_{i=p,q} \frac{1}{(p_i + k)^2} (\,\bar{p}_i k^+ -\bar{k} p^+_i\, )\,\delta^4(p_i+k+l)\, \mathcal{M}_n \,, \nn\\
&\sim&  g\,\sum_{i=p,q} \frac{1}{\langle p_i k \rangle [p_i k ]} \left(\,{[p_i k]\,\sqrt{p^+_i k^+}} \right)\,\delta^4(p_i+k+l)\, \mathcal{M}_n \,, \nn\\
&\sim&  g\,\left(\sum_{i=p,q} \frac{k^+}{\langle p_i k \rangle} \,\sqrt{\frac{p^+_i}{ k^+}} \right) \,\delta^4(p_i+k+l)\, \mathcal{M}_n \,. 
\eea
So the soft factor reads
\bea
\label{SF1}
s_{\mathcal{N}=4} = \sum_{i=p,q} \frac{k^+}{\langle p_i k \rangle} \,\sqrt{\frac{p^+_i}{ k^+}} = \frac{\langle p q\rangle}{\langle p k\rangle \langle k q\rangle}\;k^+\,.
\eea
\ndt The additional factor of $k^+$ (relative to the results in~\cite{ZWL} for example) may be traced back to the extra factors of $\partial^+$ which occur on the right hand side of (\ref {orig}) - like the $\frac{1}{\partial^+}$ in front of the first gauge field. 
\vskip 0.1cm
\ndt In appendix A, we demonstrate how the soft factor obtained above reduces correctly to the appropriate factor for gluon emission~\cite{Weinberg:1965nx} when all fermion and scalar fields are set to zero.  

\subsection{Simplified cubic vertex}
\ndt In momentum space, both cubic Lagrangians have the following basic structure (\ref{Sym})
\bea
gf^{abc}\, \frac{1}{(p^{+}+ l^{+})^2} \, (\,{p} l^+ -{l} p^+\, ) \,\tilde{{\phi}}^a(q) \tilde{\bar{\phi}}^b(p) \tilde{\bar{\phi}}^c(l) \,\delta^4(p+q+l) +c.c.
\eea
\ndt The momentum conserving delta function $\delta^4(p+q+l)$ implies that
\begin{eqnarray}
&&\langle lq\rangle =\sqrt{\frac{2}{q^+ l^{+}}}\left(pl^{+}-l p^{+}\right)=\frac{\sqrt{p^{+}}}{\sqrt{-\left(p^{+}+l^{+}\right)}}\langle pl\rangle \,,\\
&&\langle qp\rangle=\sqrt{\frac{2}{p^+ q^{+}}}\left(pl^{+}-l p^{+}\right)=\frac{\sqrt{l^{+}}}{\sqrt{-\left(p^{+}+l^{+}\right)}}\langle pl\rangle \, .
\end{eqnarray}
\ndt The cubic vertex is then
\bea
\mathcal{V}_3&=&\frac{\left(pl^+ - lp^+\right)}{(p^+ + l^+)^{\,2}}=\left[\left(\frac{pl^+ - lp^+}{p^+l^+}\right)\,( p^+ + l^+)\right]\frac{(p^+l^+)}{( p^+ + l^+)^{3}}\,,\nn\\
&=&\displaystyle \frac{\langle p l \rangle^{3}}{\langle lq \rangle \langle qp\rangle}\frac{p^+l^+}{(p^++l^+)^{3}}\ .
\eea

\subsection{Quartic vertex}
\ndt The quartic vertex is obtained using the inverse soft method. At the level of the Lagrangian, there are two types of structures possible, at quartic order, because of color ordering
\bea 
\mathcal{L}^{MHV}_{4} \sim f^{abc}\,f^{ade}\, \mathcal{V}^{I}_{4}\; \bar{\phi}^b \bar{\phi}^c \phi^d \phi^e  + f^{abc}\,f^{ade}\, \mathcal{V}^{II}_{4}\; \bar{\phi}^b \phi^c   \bar{\phi}^d \phi^e \,.
\eea
\ndt Direct multiplication of the cubic vertex (taking into account permuations of the negative helicity legs) by the soft factor produces both types of quartic vertices
\bea
&&\mathcal{V}^{I}_{4}= \frac{\langle p l \rangle^{4}}{ \langle pl\rangle \langle lq \rangle \langle q k \rangle \langle k p \rangle}\frac{p^+l^+\, k^+}{(p^++l^+)^{3}}\,,  \nn \\
&&\mathcal{V}^{II}_{4}=  \frac{\langle p q \rangle^{4}}{ \langle pl\rangle \langle lq \rangle \langle q k \rangle \langle k p \rangle}\frac{p^+q^+\, k^+}{(p^++q^+)^{3}}\,.\nn
\eea

\ndt We derive below all the component amplitudes for the $\mathcal{N}=4$ theory, so we have independent checks of this light-cone adaptation of the inverse soft method. 

\subsubsection{Checks of the result}
\ndt The $\mathcal{N}=4$ superfield in momentum space can be written as
\bea
\label{phi_0}
\phi(p)&=& e^{\frac{i}{\sqrt{2}}(\theta \bar{\theta})\,p^+} \Bigg\{\,\frac{1}{p^+}\,A(p) + \frac{i}{\sqrt{2}} \theta^m \theta^n \bar{C}_{mn}(p) + \frac{1}{12} \theta^m\theta^n\theta^p\theta^q \epsilon_{mnpq}\, p^+ \bar{A}(p) \nn\\
&&+ \frac{i}{p^+} \theta^m \bar{\chi}_m (p) + \frac{\sqrt{2}}{6} \theta^m\theta^n\theta^p \epsilon_{mnpq} \chi^q (p)\,\Bigg\} \,,
\eea
Setting scalars and fermions to zero in  (\ref{phi_0})
\bea
\label{S_0}
\phi(p)&=& e^{\frac{i}{\sqrt{2}}(\theta \bar{\theta})\,p^+} \Bigg\{\,\frac{1}{p^+}\,A(p) + \frac{1}{12} \theta^m\theta^n\theta^p\theta^q \epsilon_{mnpq}\, p^+ \bar{A}(p)\,\Bigg\} \,,
\eea
\ndt The type-$I$ quartic Lagrangian reads
\bea
\label{q_0}
\mathcal{L}^{I}_{4} = \int\!\! d^4\theta \, d^4 \bar{\theta}\;\frac{\langle p l \rangle^{4}}{ \langle pl\rangle \langle lq \rangle \langle q k \rangle \langle k p \rangle}\frac{p^+l^+\, k^+}{(p^++l^+)^{3}}\,\bar{\phi}(p)\bar{\phi}(l) \phi(q) \phi(k) \,.
\eea
\ndt Substituting (\ref{S_0}) in (\ref{q_0}) produces sixteen terms. Of these sixteen terms, only five survive Grassmann integration. These five terms are different permutations of the $4$-gluon amplitude. For example, one of these terms reads
\bea
 {M}^{MHV}_{4}(p^-\,l^-\, q^+\,k^+)= \frac{\langle p l \rangle^{4}}{ \langle pl\rangle \langle lq \rangle \langle q k \rangle \langle k p \rangle}   \,.
\eea
\ndt Using the above method, the type-$II$ vertex also gives the $4$-gluon amplitude with different permutations of negative helicity legs. \\

\ndt Gluon amplitudes with the negative helicity gluons in different positions can also be calculated using the following ward identity
\bea
M_n[p^+ l^+,......,r^- s^-,....,n^+] = \frac{\langle r s \rangle^{4}}{\langle p l \rangle^{4}}\; M_n[p^- l^- q^+ k^+,....,n^+]\,.
\eea
\\
\ndt All the other component amplitudes in $\mathcal{N}=4$ super Yang-Mills theory may then  be obtained using supersymmetric Ward identities~\cite{EH} as explained in appendix B.  
\vskip 1cm

\section{$\mathcal{N}=8$ supergravity in light-cone superspace}
We now move to the $\mathcal N=8$ theory. $\mathcal{N}=8$ superspace is spanned by the Grassmann variables $\theta^m$ and $\bar{\theta}_m (\,m=1 .... 8\,$ are $SU(8)$ indices). All $256$ physical degrees of freedom may be described by a single complex superfield. 
\bea
\label{SSG}
\phi(y)& =& \frac{1}{\del^{+\,2}}\,h(y) +i \theta^m\,\frac{1}{\del^{+\,2}}\,\bar{\psi}_m (y) + \frac{i}{2} \theta^m \theta^n \,\frac{1}{\del^{+}}\, \bar{A}_{mn}(y) \nn\\
&-& \frac{1}{3!}  \theta^m \theta^n  \theta^p \,\frac{1}{\del^{+}}\, \bar{\chi}_{mnp}(y) - \frac{1}{4!} \theta^m \theta^n  \theta^p  \theta^q \bar{C}_{mnpq} (y) \nn\\
&+& \frac{1}{5!} \theta^m \theta^n  \theta^p \theta^q \theta^r \epsilon_{mnpqrstu}\, \chi^{stu}(y) \nn\\
&+&  \frac{1}{6!} \theta^m \theta^n  \theta^p \theta^q \theta^r \theta^s \epsilon_{mnpqrstu}\, \del^+ A^{tu} (y) \nn\\
&+&  \frac{1}{7!} \theta^m \theta^n  \theta^p \theta^q \theta^r \theta^s \theta^t \epsilon_{mnpqrstu}\, \del^+ \psi^{u} (y) \nn\\
&+&  \frac{4}{8!} \theta^m \theta^n  \theta^p \theta^q \theta^r \theta^s  \theta^t \theta^u \epsilon_{mnpqrstu}\, \del^{+\,2}\bar{h} (y)\,,
\eea
where $h$ and $\bar h$ represent the graviton, ${\bar\psi}\,,\,\psi$ the spin-$\frac{3}{2}$ gravitinos, $\bar{A}_{mn}\,,\,A^{tu}$ the gauge fields,  $\bar{\chi}\,,\,\chi$ the gauginos and $\bar{C}_{mnpq}$ the $70$ real scalars. \\

\ndt The superfield $\phi$ and its complex conjugate $\bar{\phi}$ satisfy 
\bea
d^m \phi(y) =0\;\;;\;\;\;\;\;\; \bar{d}_n \bar{\phi}(y) =0 \;\; ; \;\;\;\;\;\;\phi = \frac{1}{4} \frac{d^8}{\del^{+\,4}}\, \bar{\phi} \,. 
\eea

\ndt The $\mathcal{N}=8$ supergravity action reads
\bea
S = \int\! d^4x\; d^8\theta\; d^8\bar{\theta}\; {\mathcal {L}}\ ,
\eea
where
\bea
{\mathcal {L}}=\bigg\{\; \frac{1}{4} \,{\bar{\phi}}\,\left( \frac{\Box}{\del^{+\,4}}\right)\, \phi + \frac{2}{3}\,\kappa \, \left(\frac{1}{\del^{+\,4}}\,{\bar{\phi}}\, \bar{\del}^2 \phi\, \del^{+\,2} \phi - \frac{1}{\del^{+\,4}}\,{\bar{\phi}}\, \bar{\del} \del^+ \phi\, \bar{\del}\del^{+} \phi  + c.c \right)\bigg\}\ .
\eea
Grassmann integration is normalised such that $\int\! d^8 \theta\,\theta^8 = 1$ and $\kappa = \sqrt{8\pi G}$. 
\vskip 0.3cm
\ndt The Lagrangian in momentum space at cubic order reads
\bea
\kappa \; \frac{1}{r^{+\,4}}\left(\, \bar{p}_i^2 k^{+\,2} - 2\bar{p}_ip_i^+\bar{k}k^+ + \bar{k}^2 p_i^{+\,2}\, \right) \tilde{\bar{\phi}}(r)\tilde{\phi}(p_i)\tilde{\phi}(k)\,\delta^4(p_i+k+r)\,.
\eea

\ndt In the soft limit
\bea
\mathcal{M}_{n+1}(p_1,p_2\cdots p_n,k) \sim \bigg\{\,\kappa \sum_{i=1}^n  \frac{1}{r^2 }\, \, (\,\bar{p_i} k^+ -\bar{k} p^+_i\, )^2\,\delta^4(p_i+k+r)\, \bigg\} \mathcal{M}_n(p_1,p_2\cdots,p_n) \,.
\eea
From momentum conservation
\bea
\mathcal{M}_{n+1}&\sim&  \kappa\,\sum_{i=1}^n\frac{1}{(p_i + k)^2} (\,\bar{p_i} k^+ -\bar{k} p^+_i\, )^2\,\delta^4(p_i+k+r)\; \mathcal{M}_n \,, \nn\\
&\sim&  \kappa\,\sum_{i=1}^n \frac{1}{\langle p_i k \rangle [p_i k ]} \left(\,{[p_i k]^2\,{p^+_i k^+}} \right)\,\delta^4(p_i+k+r)\; \mathcal{M}_n \,, \nn\\
&\sim&  \kappa\,\sum_{i=1}^n \frac{k^{+\,2}}{\langle p_i k \rangle} \,\left(\frac{[p_i k]\langle p_i r\rangle^2}{\langle r k\rangle^2}\right)\,\delta^4(p_i+k+r) \; \mathcal{M}_n \,.
\eea
So the soft factor is
\bea
s_{\mathcal{N}=8} = \sum_{i=1}^n\frac{[p_i k]}{\langle p_i k\rangle}\frac{\langle p_i r\rangle^2}{\langle r k\rangle^2} k^{+\,2}=\sum_{i=1}^n\frac{[p_i k]}{\langle p_i k\rangle}\frac{\langle p_ir\rangle\langle sp_i\rangle}{\langle rk\rangle\langle sk\rangle}\, k^{+\,2}\, ,
\eea
clearly a double-copy of the $\mathcal{N} = 4$ soft factor (\ref{SF1}). \\

\vskip 0.5cm

\subsection{Construction of the quartic vertices}

\ndt To construct the $\mathcal{N} = 8$ quartic vertex, we use the inverse soft method. The soft factor is
\bea
s_{\mathcal{N}=8} \;=\;\sum_{i=1}^3 \frac{[p_i k]}{\langle p_i k\rangle}\frac{\langle p_ir\rangle\langle sp_i\rangle}{\langle rk\rangle\langle sk\rangle}\,k^{+\,2} \;=\;\Bigg\{ \frac{[p k]}{\langle p k\rangle}\frac{\langle pr\rangle\langle sp\rangle}{\langle rk\rangle\langle sk\rangle}+\frac{[q k]}{\langle q k\rangle}\frac{\langle qr\rangle\langle sq\rangle}{\langle rk\rangle\langle sk\rangle}+ \frac{[l k]}{\langle l k\rangle}\frac{\langle lr\rangle\langle sl\rangle}{\langle rk\rangle\langle sk\rangle}\Bigg\}\,k^{+\,2}\,. \nn
\eea
\ndt Simplifying this using the Schouten identity and momentum conservation, we obtain
\bea
s_{\mathcal{N}=8} =\frac{[p k]}{\langle p k\rangle}\frac{\langle pl\rangle\langle pq\rangle}{\langle lk\rangle\langle kq\rangle}\,k^{+\,2}\,.
\eea
\ndt The momentum-space cubic vertex in $\mathcal{N}=8$ supergravity is
\bea
\mathcal{V}_{3}^{\mathcal{N}=8}= \frac{\langle p l \rangle^{6}}{{\langle lq \rangle}^2 {\langle qp\rangle}^2}\,\frac{p^{+\,2}\,l^{+\,2}\,q^{+\,2}}{(p^++l^+)^{8}}\,.
\eea
\ndt  So the quartic vertex reads
\bea
\label{SG4}
&&\mathcal{V}_{4}^{\mathcal{N}=8} = \displaystyle \left[\frac{\langle p l \rangle^{6}}{{\langle lq \rangle}^2 {\langle qp\rangle}^2}\,\frac{p^{+\,2}\,l^{+\,2}\,q^{+\,2}}{(p^++l^+)^{8}}\right] \times \frac{[p k]}{\langle p k\rangle}\frac{\langle p l\rangle\langle p q\rangle}{\langle l k\rangle\langle k q\rangle} k^{+\,2}\,,\nn \\ \nn\\
&&\hspace{0.7 cm}= \frac{\langle p l \rangle^{8}\,[pl]}{ \langle pl\rangle  \langle pq \rangle \langle lq\rangle   \langle pk \rangle \langle l k \rangle\langle kq\rangle^2}\,\frac{p^{+\,2}\,l^{+\,2}\,q^{+\,2}\, k^{+\,2}}{(p^++l^+)^{4}\,(q^++k^+)^{4}}\,. 
\eea
\ndt The final step involves rendering the result above symmetric under the exchange of legs, so we find 
\bea
\label{53}
&& \mathcal{V}_{4}^{\mathcal{N}=8} = \frac{\langle p l \rangle^{8}\,[pl]}{ \langle pl\rangle  \langle pq \rangle \langle lq\rangle   \langle pk \rangle \langle l k \rangle\langle kq\rangle^2}\,\Bigg\{\frac{p^{+\,2}\,l^{+\,2}\,q^{+\,2}\, k^{+\,2}}{(p^++l^+)^{4}\,(q^++k^+)^{4}} \nn\\ \nn\\
&& \hspace{2 cm}+ \frac{p^{+\,2}\,l^{+\,2}\,q^{+\,2}\, k^{+\,2}}{(p^++q^+)^{4}\,(l^++k^+)^{4}}+  \frac{p^{+\,2}\,l^{+\,2}\,q^{+\,2}\, k^{+\,2}}{(p^++k^+)^{4}\,(l^++q^+)^{4}}\Bigg\}\,. 
\eea
\ndt Similar to the case of pure gravity, in the light-cone gauge, extra terms when moving from amplitude structures to vertices - off-shell structures proportional to $p^2$ - may be shifted to higher orders in the coupling constant by suitably chosen field redefinitions. 
\vskip 0.3cm
\ndt Some checks of the result in (\ref {53}) are presented in appendix C.

\begin{center}
* ~ * ~ *
\end{center}

\ndt The possible perturbative finiteness of $\mathcal{N} = 8$ supergravity in $d = 4$ is an intriguing issue. The theory is maximally supersymmetric and hence expected to be better behaved in the ultra-violet (perturbatively) than pure gravity. Explicit calculations support this expectation and show that the theory does indeed exhibit less divergent behavior than would be naively believed. An all-order explicit proof of perturbative finiteness for $\mathcal{N} = 4$ Yang-Mills exists only in light-cone superspace~\cite{SM, BLN2}. We therefore expect that new advances in the formulation of the $\mathcal N=8$ model in this superspace are likely to allow for a more detailed finiteness analysis. Clearly an all-order proof will need vertices to all orders - however, the results of this paper will permit for a more modest task: analyze the finiteness properties of all supergraphs arising from cubic and quartic interaction vertices. This will be a much easier task to accomplish now because the existing result for the superspace quartic interaction vertices runs into hundreds of terms~\cite{ABHS}.
\\

\section*{Acknowledgments}
 The work of SA is partially supported by a MATRICS grant - MTR/2020/000073 - of SERB. SP acknowledges support from the Prime Minister’s Research Fellowship (PMRF) of the Government of India.

\pagebreak
\appendix
\section*{Appendix}
\vskip 0.5cm

\section{\!\!\!: Soft factor for gluon emission}
The $\mathcal{N}=4$ superfield in momentum space is
\bea
\label{phi}
\phi(p)&=& e^{\frac{i}{\sqrt{2}}(\theta \bar{\theta})\,p^+} \Bigg\{\,\frac{1}{p^+}\,A(p) + \frac{i}{\sqrt{2}} \theta^m \theta^n \bar{C}_{mn}(p) + \frac{1}{12} \theta^m\theta^n\theta^p\theta^q \epsilon_{mnpq}\, p^+ \bar{A}(p) \nn\\
&&+ \frac{i}{p^+} \theta^m \bar{\chi}_m (p) + \frac{\sqrt{2}}{6} \theta^m\theta^n\theta^p \epsilon_{mnpq} \chi^q (p)\,\Bigg\}\ .
\eea
Setting all scalars and fermions to zero, we are left with
\bea
\label{S1}
\phi(p)&=& e^{\frac{i}{\sqrt{2}}(\theta \bar{\theta})\,p^+} \Bigg\{\,\frac{1}{p^+}\,A(p) + \frac{1}{12} \theta^m\theta^n\theta^p\theta^q \epsilon_{mnpq}\, p^+ \bar{A}(p)\,\Bigg\}\ .
\eea
\ndt The cubic Lagrangian for $\mathcal{N}=4$ Yang-Mills in momentum space reads
\bea
\label{C1}
gf^{abc}\,\int\! d^4 \theta\, d^4 \bar{\theta}\, \frac{1}{l^{+\,2}} \, (\,\bar{p} k^+ -\bar{k} p^+\, ) \,{\bar{\phi}}^a(l) {\phi}^b(p) {\phi}^c(k) \,\delta^4(p+k+l) \,.
\eea
We substitute (\ref{S1}) in (\ref{C1}) and integrate over the Grassmann variables to obtain
\bea
  gf^{abc}\, \frac{p^+ + k^+}{p^{+}\,k^+} \, (\,\bar{p} k^+ -\bar{k} p^+\, ) \,{\bar{A}}^a(l) {A}^b(p) {A}^c(k) \,\delta^4(p+k+l) \,.
\eea
In the soft limit, the amplitude for soft gluon emission is then
\bea
 M_{n+1}\!\!\!\!\!\!\!\!&&\approx M_{n}\,\sum_{i=p,q}\frac{g}{(p_i\pm k)^2}\left(\bar p_i-p_i^+\frac{\bar k}{k^+}\right)\times\frac{p_i^++k^+}{p_i^+}\delta(p_i+l+k)\,, \nn\\
&& \sim M_n\underbrace{\,\sum_{i=p,q}\frac{gq_i}{(p_i\pm k)^2}\left(\bar p_i-p_i^+\frac{\bar k}{k^+}\right)}_{s_{YM}}\times\left(1+\frac{k^+}{p_i^+}\right)\delta(p_i+l+k)\ ,
\eea
so the soft factor associated with gluon emission is
\bea
s_{YM} =  \sum_{i=p,q}\frac{1}{\langle p_ik \rangle} \,\sqrt{\frac{p_i^+}{ k^+}} = \frac{\langle p q\rangle}{\langle p k\rangle \langle k q\rangle}\,.
\eea

\newpage

\section{\!\!\!: Component amplitudes of $\mathcal{N}=4$ super Yang-Mills}
We present independent checks of the $\mathcal{N}=4$ super Yang-Mills quartic vertex constructed using the inverse soft method in this section.
We first compute the component amplitudes using on-shell superspace and then compare them to the light-cone superspace results.
\subsection{On-shell superspace}
\ndt The $\mathcal{N}=4$ on-shell chiral superfield is
\bea
\Omega = g^+ + \eta_A \lambda^A - \frac{1}{2!}\eta_A \eta_B S^{AB} - \frac{1}{3!} \eta_A \eta_B \eta_C \lambda^{ABC} + \eta_1\eta_2 \eta_3 \eta_4 g^- \,,
\eea 
where $A,B,... = 1,2,3,4$ are labels of the global $SU(4)$ R-symmetry and $\eta$ is the on-shell Grassmann variable. \\

\ndt The tree-level MHV superamplitude of $\mathcal{N}=4$ super Yang-Mills is 
\bea
\label{A}
\mathcal{M}_n^{MHV}[123....n] = \frac{\delta^8(\tilde{Q})}{\langle 12 \rangle\langle 23 \rangle......\langle n1 \rangle}\,,
\eea
where the Grassmann delta function is defined as 
\bea
\delta^8(\tilde{Q}) = \frac{1}{2^4} \prod_{A=1}^{4} \sum_{i,j=1}^n \langle ij \rangle \eta_{iA} \eta_{jA}\,.
\eea
We can derive the component amplitudes from the MHV superamplitude as follows, \\
1) Expand the MHV superamplitude of $\mathcal{N}=4$ in Grassmann variables like the superfield. \\
2) Expand the Grassmann delta function in terms of Grassmann variables.\\
3) Compare the coefficients for the corresponding component amplitudes.\\

\ndt We can summarise all the component amplitudes in the tabular form as
{\renewcommand{\arraystretch}{2}%
\begin{center}
\begin{tabular}{ |c|c||c|c|c| } 
\hline
Amplitude & Relative factor & Amplitude & Relative factor \\
\hline \hline  
$g^- g^- g^+ g^+$ & $\langle 12 \rangle^4$ & $\lambda^{-}_{1}\,\lambda^{-}_{2}\,\lambda^{2\,+}\,\lambda^{1\,+}$& $-\langle 12 \rangle^2 \langle 13 \rangle\langle 24 \rangle$ \\ 
\hline
$g^-\,\lambda^{123}\,\lambda^{4}\,g^{+}$ & $\langle 12 \rangle^3 \langle 13 \rangle$ & $S^{12}\,\lambda^{-}\,S^{34}\,\lambda^{+}$& $\langle 12 \rangle \langle 23 \rangle^2 \langle 14 \rangle$ \\ 
\hline
$g^-\,S^{12}\,S^{34}\,g^{+}$ & $\langle 12 \rangle^2 \langle 13 \rangle^2$ & $\lambda^{-}S^{13}\,S^{14}\,\,\lambda^{+}$& $ \langle 12 \rangle\langle 13 \rangle \langle 23 \rangle \langle 34 \rangle$ \\ 
\hline
$\lambda^{234} \lambda^{134} S^{12} g^+$ & $\langle 12 \rangle^2 \langle 13 \rangle\langle 23 \rangle$ & $S^{12}\,S^{34}\,S^{34}\,S^{12}$& $ \langle 14 \rangle^2\,\langle 23 \rangle^2$ \\ 
\hline
$\lambda^{234}\,\lambda^{234}\,\lambda^{1}\,\lambda^{1}$ & $-\langle 12 \rangle^3\,\langle 34 \rangle$ & $S^{12}\,S^{23}\,S^{34}\,S^{41}$& $ \langle 12 \rangle\,\langle 23 \rangle \langle 34 \rangle\,\langle 41 \rangle$ \\ 
\hline
$\lambda^{-}_{1}\,\lambda^{-}_{2}\,\lambda^{1\,+}\,\lambda^{2\,+}$ & $-\langle 12 \rangle^2 \langle 23 \rangle\langle 41 \rangle$ & & \\ 
\hline
\end{tabular}
\end{center}}
\ndt The first entry $g^- g^- g^+ g^+$ corresponds to $M_4[1^-2^-3^+4^+]$. The common denominator factor for the amplitudes is $\langle 12 \rangle\,\langle 23 \rangle \langle 34 \rangle\,\langle 41 \rangle$.

\subsection*{Ward Identities}
\ndt Supersymmetry Ward identities of relevance to our discussions include
\bea
&& M_4[g^- \lambda^{123} \lambda^4 g^+] = \frac{\langle 13 \rangle}{\langle 12 \rangle}\,M_4[g^- g^- g^+g^+]\,. \nn\\
&& M_4[g^- S^{12} S^{34} g^+] = \frac{\langle 13 \rangle^2}{\langle 12 \rangle^2}\,M_4[g^- g^- g^+g^+]\,. \nn
\eea
We can obtain Ward identities by taking the ratio of any two amplitudes from the table above.

\vskip 0.3cm

\subsection{Light-cone superspace}
\ndt The $\mathcal{N}=4$ super Yang-Mills cubic Lagrangian reads as
\bea
\label{C1}
\mathcal{L}=gf^{abc}\, \frac{1}{(p^{+}+ l^{+})^2} \, (\,{p} l^+ -{l} p^+\, ) \,\tilde{{\phi}}^a(q) \tilde{\bar{\phi}}^b(p) \tilde{\bar{\phi}}^c(l) \,\delta^4(p+q+l) +c.c. \,.
\eea
\ndt We express the cubic vertex in three equivalent forms using momentum conservation 
\bea
\label{cubic}
\mathcal{V}_3&=& \frac{\langle p l \rangle^{3}}{\langle lq \rangle \langle qp\rangle}\frac{p^+l^+}{(p^++l^+)^{3}}\,.\nn\\
\mathcal{V}'_3&=& {\langle p l \rangle}\frac{\sqrt{p^+l^+}}{(p^++l^+)^{2}}\,. \nn\\
\mathcal{V}''_3&=& \frac{\langle l q\rangle \langle pq \rangle}{ \langle pl\rangle}\frac{1}{q^+}\,.
\eea
\ndt The $\mathcal{N}=4$ super Yang-Mills soft factor is
\bea
\label{SF1}
s_{\mathcal{N}=4} = \sum_{i=p,q} \frac{k^+}{\langle p_i k \rangle} \,\sqrt{\frac{p^+_i}{ k^+}} = \frac{\langle p q\rangle}{\langle p k\rangle \langle k q\rangle}\;k^+\,.
\eea
\ndt The quartic vertex can be constructed using the inverse soft method. At the level of Lagrangian, two types of structure are possible for quartic vertex because of color ordering. In the subsection below, we will work with the first type of vertex to derive the component amplitudes.\\
\subsubsection{ Component amplitudes}
\ndt The recipe for computing the component amplitudes is as follows: \\
(i) Start with the appropriate quartic Lagrangian. \\
(ii) Substitute the expression for the superfield in the Lagrangian. \\
(ii) Integrate out the Grassmann variables and read off the component amplitudes\\

\subsection*{\small 2-fermion--scalar--gluon amplitude}

\ndt Take $\mathcal{V}'_{3}$ from (\ref{cubic}) and construct quartic vertex using inverse soft method
\bea
\label{q4}
\mathcal{L}_4 &=& \int\!\! d^4\theta \, d^4 \bar{\theta}\;\langle p l \rangle \frac{\sqrt{p^+l^+}}{(p^+ + l^+)^{\,2}}  \frac{\langle p q\rangle}{\langle p k\rangle \langle k q\rangle}\;k^+\;\bar{\phi}(p)\bar{\phi}(l) \phi(q) \phi(k) \,.
\eea

\ndt Substitute (\ref{phi}) in (\ref{q4}) and integrate out the Grassmann variables. The 2-fermion--scalar--gluon amplitude is
\bea
{M}^{MHV}_{4}\left[\chi^m(p)\,\chi^n(l)\,\bar{C}_{mn}(q)\, A(k)\right] = \frac{\langle p l \rangle\, \langle pq \rangle}{   \langle q k \rangle \langle k p \rangle}\,.
\eea
\ndt The above amplitude matches with the one derived using the on-shell superspace.

\subsection*{\small 4-fermion amplitude}
\ndt The quartic Lagrangian from (\ref{q4}) reads as
\bea
\label{q5}
\mathcal{L}_{4} =\int\!\! d^4\theta \, d^4 \bar{\theta}\; \langle p l \rangle \frac{\sqrt{p^+l^+}}{(p^+ + l^+)^{\,2}}  \frac{\langle p q\rangle}{\langle p k\rangle \langle k q\rangle}\;k^+\,\bar{\phi}(p)\bar{\phi}(l) \phi(q) \phi(k)\ .
\eea
Setting gluons and scalars to zero in the superfield (\ref{phi})
\bea
\label{S5}
\phi(p)= e^{\frac{i}{\sqrt{2}}(\theta \bar{\theta})\,p^+} \Bigg\{ \frac{i}{p^+} \theta^m \bar{\chi}_m (p) + \frac{\sqrt{2}}{6} \theta^m\theta^n\theta^p \epsilon_{mnpq} \chi^q (p)\,\Bigg\}\ .
\eea
Substituting (\ref{S5}) in (\ref{q5}), we get
\bea
\label{4f}
{M}^{MHV}_{4}\left[\chi^m(p)\,\chi^n(l)\,\bar{\chi}_{m}(q)\,\bar{\chi}_{n}(k)\,\right] = \frac{\langle p l \rangle^2 \langle p q \rangle}{   \langle q k \rangle \langle k p \rangle \langle ql \rangle } \sqrt{\frac{k^+p^+}{q^{+\,2}}}.
\eea

\subsection*{\small 2-fermion--2-scalar amplitude}
\ndt Using $\mathcal{V}''_{3}$, we obtain the quartic vertex for $\mathcal{N}=4$ super Yang-Mills 
\bea
\label{q6}
\mathcal{L}_{4} =\int\!\! d^4\theta \, d^4 \bar{\theta}\;\frac{\langle lq \rangle\, \langle pq \rangle}{  \langle pl \rangle}\,\frac{1}{q^+}\,\frac{\langle p q\rangle}{\langle p k\rangle \langle k q\rangle}\,k^+\;\bar{\phi}(p)\bar{\phi}(l) \phi(q) \phi(k) \,.
\eea
After setting gluons to zero in the superfield (\ref{phi}), substitute it in (\ref{q6}) and integrate out all the Grassmann variables. We obtain
\bea
\label{2s2f}
{M}^{MHV}_{4}\left[C^{mn}(p)\,\chi^a(l)\bar{C}_{mn}(q)\,\bar{\chi}_{a}(k)\,\right] = \frac{\langle pl \rangle \langle pq \rangle}{   \langle qk \rangle \langle kp \rangle } \sqrt{\frac{p^+ k^+}{q^{+\,2}}}.
\eea
\ndt Take the $4$-fermion amplitude from (\ref{4f}) and divide it by (\ref{2s2f}),
\bea
M_4[\chi^m\,\chi^n\,\bar{\chi}_m\,\bar{\chi}_n] = \frac{ \langle pl \rangle}{ \langle ql \rangle}\, {M_4[C^{mn}\,\chi^a\bar{C}_{mn}\,\bar{\chi}_{a}}]\,.
\eea
\ndt We get one of the supersymmetric Ward identities, which implies that the derived component amplitudes are correct. We have worked out in detail only a few of the component amplitudes here. All the other amplitudes can be similarly derived using this method.

\section{\!\!\!: Checks of the result for $\mathcal{N}=8$ supergravity}

\ndt The Lagrangian, at quartic order, constructed using the inverse soft approach is
\bea
\label{SG0}
\mathcal{L}^{MHV}_{4}= \frac{\langle p l \rangle^{8}\,[pl]}{ \langle pl\rangle  \langle pq \rangle \langle lq\rangle   \langle pk \rangle \langle l k \rangle\langle kq\rangle^2}\,\frac{p^{+\,2}\,l^{+\,2}\,q^{+\,2}\, k^{+\,2}}{(p^++l^+)^{4}\,(q^++k^+)^{4}}\,\bar{\phi}(p)\bar{\phi}(l) \phi(q) \phi(k). 
\eea
\ndt To obtain the four-point pure gravity amplitude, we start with the $\mathcal{N}=8$ superfield (\ref{SSG}) and set all fields, other than the graviton, to zero. 
\bea
\label{phi_01}
\phi(p)&=& e^{\frac{i}{\sqrt{2}}(\theta \bar{\theta})\,p^+} \Bigg\{\,\frac{1}{p^{+\,2}}\,h(p)\,+\, \frac{4}{8!}\, \theta^m \theta^n  \theta^p \theta^q \theta^r \theta^s  \theta^t \theta^u \epsilon_{mnpqrstu}\, p^{+\,2}\bar{h} (p)  \Bigg\}\,.
\eea
We have sixteen terms after substituting the superfield (\ref{phi_01}) in  (\ref{SG0}). Only five of those sixteen terms survive Grassmann integration. These five terms are different permutations of the four-graviton amplitude, for example 
\bea
{M}^{G}_{4}(p^-\,l^-\,q^+\,k^+)= \frac{\langle p l \rangle^{8}\,[pl]}{ \langle pl\rangle  \langle pq \rangle \langle lq\rangle   \langle pk \rangle \langle l k \rangle\langle kq\rangle^2}.
\eea
\ndt All other $\mathcal{N}=8$ supergravity component amplitudes may be obtained similarly.\\

\ndt Another check of the result in (\ref{SG0}) is the following. Project out the component amplitude ${M}^{MHV}_{4}(C^{mnpq}\,C^{rstu}\,h^-\,h^+)$ from (\ref{SG0}). This reads
\bea
{M}^{MHV}_{4}(C^{mnpq}\,C^{rstu}\,h^-\,h^+)=  \frac{\langle p q \rangle^{4}\,\langle ql \rangle^{4}\,[pl]}{ \langle pl\rangle  \langle pq \rangle \langle lq\rangle   \langle pk \rangle \langle l k \rangle\langle kq\rangle^2}\,.
\eea
\ndt If the momentum of the external scalar is taken soft, then this amplitude vanishes 
\bea
\lim_{p\to 0}\;{M}^{MHV}_{4}(C^{mnpq}\,C^{rstu}\,h^-\,h^+) =0\ ,
\eea
matching expectations from the literature~\cite{EH}.
\pagebreak

\end{document}